%% file: main.tex
\documentclass{article}

\usepackage{fullpage}
\usepackage{amsmath,amsfonts}
\usepackage{times}
\usepackage{amsthm}
\usepackage{graphicx}
\usepackage{enumitem}
\usepackage{color}   
\usepackage{framed}
\usepackage{hyperref}
\hypersetup{
    colorlinks=true, 
    linktoc=all,     
    linkcolor=blue,  
    citecolor=red,
}

\newcommand{\cnote}[1]{}
\newcommand{\mnote}[1]{}
\newcommand{\anote}[1]{}

\newcommand{\remove}[1]{}

\newcommand{\inp}{\mathsf{Inp}}

\newcommand{\Dec}{\mathsf{Dec}}

\renewcommand{\paragraph}[1]{\medskip\noindent{\bf #1}}

\newcommand{\dec}{\mathsf{Decode}}

\newcommand{\ccc}{\mathrm {CC}}

\newcommand{\coin}{{\scriptscriptstyle\mathrm{COIN}}}
\newcommand{\ot}{{\scriptscriptstyle\mathrm{OT}}}

\newcommand{\ole}{{\scriptscriptstyle\mathrm{OLE}}}

\newcommand{\comm}{{\scriptscriptstyle\mathrm{COM}}}

\newtheorem{thm}{Theorem}
\newcommand{\BT}{\begin{thm}}   \newcommand{\ET}{\end{thm}}
\newtheorem{dfn}{Definition}      %
\newcommand{\BD}{\begin{dfn}}   \newcommand{\ED}{\end{dfn}}
\newtheorem{corr}[thm]{Corollary}      %
\newcommand{\BCR}{\begin{corr}} \newcommand{\ECR}{\end{corr}}
\newtheorem{Ithm}{Theorem}[section]     
\newcommand{\BIT}{\begin{Ithm}}   \newcommand{\EIT}{\end{Ithm}}
\newtheorem{lem}{Lemma}[section]  
\newcommand{\BL}{\begin{lem}}   \newcommand{\EL}{\end{lem}}
\newtheorem{prop}[lem]{Proposition}
\newcommand{\BP}{\begin{prop}}   \newcommand{\EP}{\end{prop}}
\newtheorem{clm}[lem]{Claim}            %
\newcommand{\BCM}{\begin{clm}}   \newcommand{\ECM}{\end{clm}}
\newtheorem{fact}[lem]{Fact}            %
\newcommand{\BF}{\begin{fact}}   \newcommand{\EF}{\end{fact}}
\newcommand{\BPF}{\begin{proof}} \newcommand {\EPF}{\end{proof}}
\newtheorem{prot}{Protocol}      
\newcommand{\BPR}{\begin{prot}}   \newcommand{\EPR}{\end{prot}}
\newenvironment{cproof}{\noindent{\bf Proof:~~}}{\hfill $\Box$}
\newcommand{\BCPF}{\begin{cproof}} \newcommand {\ECPF}{\end{cproof}}

\newtheorem{remark}[lem]{Remark}            %
\newcommand{\BR}{\begin{remark}}   \newcommand{\ER}{\end{remark}}

\newcommand{\BDE}{\begin{description}}
\newcommand{\EDE}{\end{description}}
\newcommand{\BE}{\begin{enumerate}}
\newcommand{\EE}{\end{enumerate}}
\newcommand{\BI}{\begin{itemize}}
\newcommand{\EI}{\end{itemize}}
\newcommand{\BEQ}{\begin{eqnarray*}}
\newcommand{\EEQ}{\end{eqnarray*}}
\def\blackslug
{\hbox{\hskip 1pt\vrule width 8pt height 8pt depth 1.5pt\hskip 1pt}}
\def\qed{\quad\blackslug\lower 8.5pt\null\par}

\newcommand{\cir}{\mathrm{C}}

\newcommand{\cA}{{\mathcal A}}

\newcommand{\cF}{{\mathcal F}}

\newcommand{\cI}{{\mathcal I}}

\newcommand{\cS}{{\mathcal S}}

\newcommand{\cZ}{{\mathcal Z}}

\newcommand{\FF}{\mathbb{F}}

\newcommand{\RS}{\mathsf{RS}}

\newenvironment{boxfig}[2]{\begin{figure}[#1]\fbox{\begin{minipage}{.95\columnwidth}
                        \vspace{0.2em}
                        \makebox[0.025\columnwidth]{}
                        \begin{minipage}{0.95\columnwidth}
            {\small{
                        #2 }}
                        \end{minipage}
                        \vspace{0.2em}
                        \end{minipage}}}{\end{figure}}

\newcommand{\pprotocol}[4]{
\begin{boxfig}{!t}{
\begin{center}
{\bf #1}
\end{center}
    #4
\vspace{0.5ex} } \caption{ #2}\label{#3}
\end{boxfig}
}

\newcommand{\protocoll}[4]{
\pprotocol{#1}{#2}{#3}{#4} }

\newcommand{\Alice}{\mathrm P_0}
\newcommand{\Bob}{\mathrm P_1}
\newcommand{\party}{\mathrm P_i}

\newcommand{\client}{\mathrm C_i}
\newcommand{\alice}{\mathrm C_0}
\newcommand{\bob}{\mathrm C_1}
\newcommand{\watch}{\mathsf{W}}

\begin{document}

\title{Oursourcing Private Machine Learning via \\Lightweight Secure Arithmetic Computation}

\author{Siddharth Garg \thanks{NYU Tandon School of Engineering. Email: {\tt{sg175@nyu.edu,ghodsi@nyu.edu}}} \and Zahra Ghodsi $^*$ \and Carmit Hazay \thanks{Bar-Ilan University. Email: {\tt{carmit.hazay@cs.biu.ac.il.}}} \and Yuval Ishai \thanks{Technion. Email: {\tt{yuvali@cs.technion.ac.il}}} \and Antonio Marcedone  \thanks{Cornell University. Email: {\tt{am2623@cornell.edu}}} \and Muthuramakrishnan Venkitasubramaniam \thanks{University of Rochester. Email: {\tt{muthuv@cs.rochester.edu}}}}
\date{}

\maketitle

\begin{abstract}
In several settings of practical interest, two parties seek to collaboratively perform inference on their private data using a public machine learning model. For instance, several hospitals might wish to share patient medical records for enhanced diagnostics and disease prediction, but may not be able to share data in the clear because of privacy concerns. In this work, we propose an {\em actively secure} protocol for outsourcing secure and private machine learning computations. Recent works on the problem have mainly focused on passively secure protocols, whose security  holds against passive (``semi-honest'') parties but may completely break down in the presence of active (``malicious'') parties who can deviate from the protocol. Secure neural networks based classification algorithms can be an seen as an instantiation of an {\em arithmetic computation} over integers. 
%
%
%
We showcase the efficiency of our protocol by applying it to real-world instances of arithmetized neural network computations, including a network trained to perform collaborative disease prediction.


\end{abstract}


\input{intro}

\newcommand{\mainbody}[1]{#1}
\newcommand{\moveappendix}[1]{}


\input{implementation}

\input{benchmark}

{\small
\bibliographystyle{alpha}
\bibliography{main}}

\appendix

\input{ips}

\input{inner}

\renewcommand{\mainbody}[1]{}
\renewcommand{\moveappendix}[1]{#1}

\end{document}

%% file: intro.tex

\section{Introduction}

In this work, we consider the problem of privacy-preserving outsourcing of machine learning computations where a set of clients $P_1,\ldots,P_m$ wish to jointly classify private inputs based on a machine learning model by outsourcing the computation to a set of cloud (compute) servers. We are interested in scenarios where the model (weights) can either be public (i.e., known to everyone) or private (i.e. provided as input by one of the clients).

More formally, we envision our protocol in three phases. In a first phase, the clients share their inputs with the cloud servers, followed by a compute phase, where the servers are tasked with applying a machine learning model jointly on the client's inputs. In the third and final phase, the servers deliver the result of the inference back to the clients.  Security here demands that classification remain private as long as there is one honest cloud server, however, correctness should hold even if all the servers are corrupted.
In the context of outsourcing privacy-preserving machine learning computations past works have considered a weaker variant of this formulation where an honest server is required to guarantee correctness \cite{MohasselZ17,LiuJLA17,MakriRSV17}. With the exception of the work by Markri et al. \cite{MakriRSV17}, all previous works only consider passive security \cite{MohasselZ17,LiuJLA17,RiaziWTS0K18,JuvekarVC18,WaghGC18} where the adversary corrupting the cloud servers is assumed to honestly follow the protocol. The EPIC system from \cite{MakriRSV17} achieves security in the face of active corruptions where the adversary is allowed to arbitrarily deviate.  
\mnote{Move this later.}
While EPIC does allow the data outsourcing party to pre-process its data via a neural network feature extractor, the network's weights must be agnostic to the classification task. This constraint limits the neural network's classification accuracy.

Providing security against active adversaries in our setting presents its own challenges beyond utilizing an active secure protocol for the underlying functionality. In more details, the parties must be ensured that the servers use the ``right'' inputs and do not abuse their valid inputs by adding to them a carefully chosen small adversarial perturbation with the aim to change the predication \cite{TabacofV16,KreukACK18}. These type of attacks can be devastating when correctness of computation is crucial to the application such as in medical diagnosis and image classification for defense applications, and only arise in the presence of active adversaries. 


\subsection{Background and Related Work}

 Deep neural networks, in general, implement multiple layers of computation.
The inputs to a layer are multiplied by a matrix of trained weights, followed by an element-wise non-linear activation function.
For classification tasks, the outputs of the final layer are passed through a special activation function, referred to as the \emph{softmax}, that returns the probabilities of the input belonging to each class. Like prior work that applies cryptographic techniques on neural network computations~\cite{gilad2016cryptonets,GhodsiGG17}, we focus on neural networks that can be represented as \emph{arithmetic circuits}. Specifically, we constrain the inputs and weights of the network to be field elements and the activation functions  
to be quadratics, i.e., the output is the square of the input. We note that networks with quadratic activations have the same representational power as networks with more commonly used activations such as the ReLU function \cite{gautier2016globally,livni2014computational}.

The study of privacy-preserving machine learning has recently witnessed prosperity due to the impressive progress in the design of practical secure multiparty computation (MPC) protocols, which mostly led to passively secure protocols thus far. We review some of these notable works. In \cite{MohasselZ17} Mohassel et al. introduce SecureML, a system for several privacy preserving machine learning training and classifications algorithms in the two-server model that run 2PC for arithmetic computation. In \cite{LiuJLA17} Liu et al. develop MiniONN, a framework for transforming an existing neural network to an oblivious neural network that protects the privacy of the model (held by a cloud) and the client's input in the predication phase. In \cite{BarnettSSSSVW17}, somewhat homomorphic encryption is used for image classification focusing, on the particular non-linear Support Vector Machines algorithm. In \cite{RiaziWTS0K18} Riazi et al. present Chameleon, a system that supports hybrid secure computation in the two-party setting that combines arithmetic computation over rings for linear operations and Yao's garbled circuits \cite{Yao86} for the non-linear computation. Chameleon provides training and classification for deep and convolutional neural networks. Juvekar et al. \cite{JuvekarVC18} extends this paradigm in GAZELLE for classifying private images using a convolutional neural network, protecting the classification phase and using fully homomorphic encryption scheme for carrying out the linear computation. This construction was later improved in \cite{MakriRSV17} that introduced EPIC, an outsourcing scheme with active security. Finally, in a recent work by Wagh et al. \cite{WaghGC18}, the authors introduce SecureNN, a tool for training and predication in the three-party and four-party settings with honest majority. Their secure training implements three types neural networks, optimizing on prior training protocols.

Concretely efficient general secure two-party (2PC) and multi-party computation (MPC) has seen a lot of progress in the last decade with a focus on boolean computations \cite{LindellP12,ShelatS13,LindellPSY15,WangRK17,KatzRW17}. However, in many applications, the computation can be more naturally described by using arithmetic operations over integers, real numbers, or other rings. For such instances of {\em secure arithmetic computation}, general techniques for securely evaluating Boolean circuits incur a very significant overhead (see, e.g., \cite{ApplebaumIK14,BallMR16,Ben-Efraim17} and references therein). For secure computation protocols that directly compute arithmetic functionalities, there have been two basic approaches: Oblivious-Transfer (OT) based \cite{Gilboa99,KellerOS16,FrederiksenPY18} and Oblivious Linear Evaluation\footnote{An OLE-based protocol employs a secure two-party protocol for computing the function $ax+b$ over $\FF$ where $a,b$ is provided as input from one party and $x$ from the other. OLE is a commonly used primitive for secure arithmetic computation, analogously to the role of oblivious transfer in secure Boolean computation~\cite{NaorP99,IshaiPS09,ApplebaumDI0Z17}.} (OLE) based \cite{DamgardPSZ12,DottlingGNNT17,GhoshNN17,KellerPR18}. Generally speaking, the OT-based approach is light on the computation but involves a higher communication cost for secure arithmetic computation over large fields.
On the other hand, the OLE-based implementations in prior works have required a specific OLE implementation and/or significant invocations of the OLE functionality per multiplication gate of the underlying arithmetic circuit. Moreover, all protocols have suffered a signficant overhead when moving from a passive secure protocol to an active one.

\subsection{Our Contribution}

In this work, we introduce a new model for securely outsourcing arithmetic computations to a set of untrusted cloud servers. We discuss an appealing use case of this application for outsourcing the classification of a (potentially) propriety neural network to two cloud servers. In our model, we provide secrecy when at least one cloud server is honest (i.e. non-colluding) and guarantee correctness even when all servers collude. 

\anote{I don't think the 2PC is a contribution of this work. What is the difference with Leviosa here?}\mnote{fixed.} Next, we design 
a concretely efficient secure two-party computation protocol for general arithmetic circuits. A generic approach to achieve our goal would be for the servers to rely on a secure arithemtic computation protocol to compute the result and a zkSNARK proof of correctness of the computation \cite{BitanskyCCT12}. However, this would be quite inefficient as the prover computations in these proof systems are typically heavy. Instead we rely on the secure arithmetic computation protocol of  \cite{MHIV18} where we show a zkSNARK proof can be attached with minimal overhead.  
In slight more detail, the work of \cite{MHIV18} provide a secure two-party arithmetic computation protocol with significant efficiency improvements over previous protocols of this type. For example, for typical arithmetic circuits (that are large but not too narrow), it yields over 15x improvement over the recent Overdrive~\cite{KellerPR18} and TinyOLE~\cite{DottlingGNNT17} protocols in terms of both communication and computation. 

A distinctive feature of this protocol is that it can make a fully modular {\em black-box} use of any {\em passively secure} implementation of oblivious linear function evaluation (OLE), which serves as a natural building block for secure arithmetic computation. This means that it can build on a variety of existing or future implementations of passively secure OLE, inheriting their security and efficiency features. Given the multitude of optimization goals, security requirements  and execution platforms, such a modular design can have major advantages. 


We benchmark our protocol against our privacy-preserving machine learning system.
As an instance of such a system, we train a two-layer arithmetic neural network that predicts the drug usage of patients based on data obtained from multiple health services providers including the patients' personal and psychological traits and their usage of other drugs. The trained arithmetic neural network achieves $85\%$ accuracy and is comparable to that of a conventional neural network with floating point weights and ReLU activations.


\section{Our Model}

In our model, we consider different players who interact in synchronous rounds via authenticated point-to-point channels and a broadcast medium. Players can be designated three different roles: input clients who provide inputs, output clients who receive outputs, and servers who may be employed for the actual computation. Our protocols are described for the specific case of two servers but can be extended to constant number of servers following \cite{IshaiPS08}. We assume two different adversary models with different security guarantees:
\begin{enumerate}
\item We consider a static rushing adversary that corrupts all but one of the input clients and all but one of the servers. In this scenario, we achieve security with abort, i.e. an adversary is restricted to modifying the inputs of corrupted input clients and can make the protocol abort but cannot violate the privacy of the inputs from uncorrupted input clients or correctness of the computation if it proceeds to completion. 
\item We consider a second scenario where the adversary is static and rushing and can corrupt all-but-one of the input clients and \emph{all} the servers. In this scenario, the adversary can modify the inputs of corrupted input clients and learn the inputs of the uncorrupted clients but not affect the correctness of the output of the computation if it proceeds to completion.
\end{enumerate}

\subsection{Defining Security}
In this section, we present the security definitions considered in this paper. We follow the standard simulation paradigm and define the real-world and ideal-world executions.

\paragraph{The real model.} Let $\pi$ be a multiparty protocol computing a deterministic functionality $f$. Following \cite{Canetti01}, we consider the protocol execution in the presence of an adversary $\cA$ that is coordinated by an environment $\cZ$ that is modeled as non-uniform probabilistic polynomial time Turing machine. At the beginning of the computation, the environment chooses and delivers the inputs to the parties. More formally, the environment sends $(1^\kappa,1^s, x_i)$ to each party $P_i$ where $\kappa$ and $s$ are respectively the computational and statistical security parameter. We denote by $\cI$ the set of corrupted parties. $\cZ$ sends $\{x_i\}_{i \in \cI}$ and $z$ to $\cA$. The the parties interact according to protocol $\pi$, where the honest parties follow the instructions as specified by $\pi$ which the corrupted parties behave arbitrarily as dictated by $\cA$. At the end of the protocol $\cA$ provides an output to $\cZ$ and $\cZ$ additionally receives the output of the honest parties. Finally $\cZ$ outputs a bit and is denoted by the random variable $\textsc{real}_{\pi,\cA,\cZ}(\kappa,s)$. 

\paragraph{The ideal model - security with abort.} In the ideal model, we consider a trusted party that implements the functionality $f$. Here too, we have an environment $\cZ$ that provides the inputs to the honest parties and the inputs of corrupted parties and $z$ to an adversary $\cS$. Execution in the ideal model proceeds as follows. Honest parties submit their input $x_i$ to the trusted party implementing $f$. Corrupted parties can send arbitrary values as instructed by the adversary $\cS$. We denote this by $x'_i$ for $i \in \cI$. The trusted party computes $f(x_1,\ldots,x_n)$ and sends it to the adversary $\cS$. The adversary then decides and instructs the functionality to deliver the output to the honest parties.  As in the real world, the adversary $\cS$ sends an output to the environment. We denote the output of the environment ideal world by  $\textsc{ideal}_{f,\cA,\cZ}(\kappa,s)$. 

\paragraph{The ideal model - verifiable computation.} This ideal model is identical to the case of security with abort, with the only exception being, the adversary $\cS$ is given the inputs of all parties (not just the inputs of the corrupted parties). In other words, in this model, the adversary learns the inputs of all parties but cannot affect the correctness of the computation. We denote the output of the environment ideal world by  $\textsc{ideal}^{comp}_{f,\cA,\cZ}(\kappa,s)$. 

\BD[Security with abort]\label{def1} Let $\pi$ be a protocol and $f$ be an $n$-party functionality . We say that $\pi$ securely computes $f$ in the presence of static  active adversaries corrupting the parties $\cI$ if for any PPT adversary $\cA$ there exists a PPT adversary $\cS$ such that for every non-uniform PPT $\cZ$ there is a negligible function $\mu(\cdot)$ such that 
$$|\Pr[\textsc{real}_{\pi,\cA,\cZ}(\kappa,s) = 1] - \Pr[\textsc{ideal}_{f,\cA,\cZ}(\kappa,s)=1]| \leq \mu(\kappa)+2^{-s}$$
\ED
\BD[Verifiable Computation]\label{def2} Let $\pi$ be a protocol and $f$ be an $n$-party functionality . We say that $\pi$ securely computes $f$ in the presence of static  active adversaries corrupting the parties $\cI$ if for any PPT adversary $\cA$ there exists a PPT adversary $\cS$ such that for every non-uniform PPT $\cZ$ there is a negligible function $\mu(\cdot)$ such that 
$$|\Pr[\textsc{real}_{\pi,\cA,\cZ}(\kappa,s) = 1] - \Pr[\textsc{ideal}^{comp}_{f,\cA,\cZ}(\kappa,s)=1]| \leq \mu(\kappa)+2^{-s}$$
\ED

\section{Our Secure Arithmetic Computation Protocol}
\anote{Can we call our protocol 2PC? Even if there are two servers, there are multiple clients and parties who get output} In this work, we design, optimize, and implement an {\em actively secure} protocol for secure arithmetic computation based on the work of \cite{MHIV18}. 
This protocol is based on the IPS-compiler, introduced by Ishai et al. \cite{IshaiPS08}, which provides a general paradigm to design secure computation protocol combining the following two ingredients: (1) an active MPC protocol for the underlying function $f$ in the honest majority setting, and (2) a passive MPC protocol of a related functionality in the dishonest majority setting. \anote{Muthu please check the next sentence, I reworded it: } The work of \cite{MHIV18} optimizes both the analysis and the building blocks of the IPS-compiler and constructs a protocol which requires roughly 4 invocations of any passive OLE implementation per multiplication gate of the circuit. A crucial ingredient in the construction that enables the throughput is the use of Reed-Solomon codes to implement a packed secret sharing scheme \cite{FranklinY92}.

\anote{This is confusing: leviosa is 2PC, there is no cloud servers:} First we recall the protocol from \cite{MHIV18} that guarantees correctness and privacy when one of the parties remains honest. This is similar to the guarantees provided in the work of Makri et al. \cite{MakriRSV17}. 
However, there is a subtle issue in their protocol that arises when outsourcing computations to malicious servers.
To achieve security against all-but-one corruptions of the servers there needs to be a way to ensure that the inputs from all the parties have been correctly included in the computation (i.e. not been tampered with). 
Without such a mechanism, naively employing an actively secure protocol could allow an adversary to alter the input (via an ``additive'' attack) of any party before the computation starts. To mitigate such an attack, they employ a simple information theoretic mac. \anote{We just do MAC(m,k) = km in the implementation. This is NOT a MAC, although I agree it is ok for our purposes. MAC(m,k1,k2)} In slight more detail, they consider a slightly modified (randomized) functionality $\cF'$ that the servers compute. Roughly speaking, the functionality takes as input MAC keys $k_x$ and $k_y$ and MACs $m_x$ and $m_y$, in addition to the inputs $x$ and $y$, and produces as output $(f(x,y),{\rm flag})$. The flag bit will be set to a random linear combination of the values $d_x=m_x - {\rm MAC}_{k_x}(x)$ and $d_y=m_y - {\rm MAC}_{k_y}(y)$. Finally, we remark that this approach can be modularly combined with any 2PC, including the work of  \cite{MakriRSV17} to protect against such an attack.

\anote{Intuitively it seems to me that if the clients release hashes of their inputs that are needed to check correctness of the computation, then the MAC and flag constructions above is not required any more. Should we stress this? Also I think we really need a description of the protocol, otherwise the main contribution of this work is this paragraph, which doesn't really say much at all.} In our protocol, we enhance this basic protocol from \cite{MHIV18} to guarantee correctness even when all cloud servers are corrupt. In other words, in our protocol, an active adversary corrupting all cloud servers can potentially violate the privacy of the inputs but will be caught if it delivers an incorrect answer. In slight more detail, our approach takes advantage of the IPS compiler to additionally attach a ``zero-knowledge'' proof with the result of the computation. Roughly speaking, the views of a random subset of the parties emulated as part of the virtual outer protocol serves as the ``proof''. This is inspired by the approach of Ishai et al. \cite{IshaiKOS07} and refined by Ames et al. \cite{AmesHIV17}. Overall this variant requires two modifications to the underlying basic protocol: (1) Each party needs to publish a commitment (that can be implemented via a publicly verifiable ``hash'' function), and (2) The cloud servers compute a proof that attests that the inputs considered in the computation correspond to the public hash values released by the clients. In the full version of the paper, we provide the complete description of the protocol and security analysis. Formally, we provide security with abort if the adversary corrupts all-but-one of the servers according to Definition~\ref{def1} and verifiable computation if all the servers are corrupted according to Definition~\ref{def2}. In this sense, we get a best of both worlds type of guarantee. 

Finally, we re-iterate  that most previous approaches for securely computing machine learning tasks have considered only passive security and the ones that consider active security cannot provide this additional verifiability feature in the presence of an active adversary corrupting all cloud servers.

Below, we benchmark the basic variant of our protocol to a concrete NN computation and leave implementing the enhanced version as future work. The details of the protocol from \cite{MHIV18} is provided (verbatim) in Appendix~\ref{app:leviosa} for immediate reference. 

%
%

%% file: implementation.tex

\remove{
\begin{figure*}[t]
   \begin{minipage}{.30\linewidth}
    \includegraphics[width=\columnwidth]{authTriple.pdf}
    \caption{Communication throughput for generating authenticated triples.}\label{fig:multitime}
  \end{minipage}
\hfill
  \begin{minipage}{.30\linewidth}
     \includegraphics[width=\columnwidth]{authTripleTime.pdf}
     \caption{ Throughput for generating authenticated triples.}\label{fig:authTriple}
  \end{minipage}
  \hfill
    \begin{minipage}{.30\linewidth}
     \includegraphics[width=\columnwidth]{oleGraph.pdf}
     \caption{ Active to Passive OLE overheads.}\label{fig:ole}
  \end{minipage}
\end{figure*}%
}

\section{Implementation And Results}\label{sec:eval}

For evaluating our proposed model, we considered the Drug Consumption dataset~\cite{fehrman2017five} obtained from the UCI repository. This dataset consists of drug usage records of $19$ drugs for $1885$ respondents. For each respondent $12$ attributes are known, including $7$ personality traits obtained from psychological surveys, as well as $5$ personal information: level of education, age, gender, country of residence and ethnicity. We are envisioning a scenario in which personality traits are available to one health center, and drug usage for a number of drugs are available to another, and the two health centers want to collaboratively predict the risk of being a drug user for another drug.
We trained a $2$ layer fully connected neural network with $20$ neurons in each layer to predict if a respondent is a user of LSD. We used $80\%$ of the data for training the the remaining for test. A quadratic function was used as the activation function in the hidden layers.
Training the network on all features including personality traits, personal information as well as drug usage record for other drugs achieves a test accuracy of $85.94\%$. It is worth noting that if each of the health centers have trained a model on the data available to them, the final accuracy would drop, with $76.92\%$ test accuracy when trained only on personality traits and personal information, and $83.29\%$ test accuracy when trained only on records for other drug usage and personal information. Using quadratic activations didn't degrade the accuracy, as networks trained with ReLU activation achieved similar accuracies ($85.68\%$, $78.25\%$ and $83.82\%$ respectively).

%% file: benchmark.tex

\paragraph{Benchmark.}
We implemented our protocol in C++ using Shoup's NTL library \cite{shoup} to perform arithmetic over finite fields. 
We relied on a recent lightweight passive OLE implementation due to Catro et al. that is based on the LWE assumption \cite{Chiraag}. 
We implemented our protocol in C++ using Shoup's NTL library \cite{shoup} to perform arithmetic over finite fields, in particular vector linear operations and discrete Fast Fourier Transforms. We chose an appropriate prime of length at least 60-bits with sufficiently many roots of unity. This is required for our packed secret sharing routine.  Moreover, we constrained our parameters to fix the degree of the secret sharing  in the outer protocol be a power of two. This is so that we achieve optimal efficiency through FFT algorithms in the finite field. We implemented symmetric encryption by xoring our plaintexts with a pseudorandom mask generated by NTL's PRG implementation on input a 256 bit seed. In addition, we used the cryptoTools library \cite{cryptotools} maintained by Peter Rindal for the commitment functionality, as well as network communication (which in turn is based on Boost Asio). We also used the libOTe library \cite{libote} for implementing the $k$-out-of-$n$ oblivious transfer functionality which is maintained by Peter.
We used the batched version of the (batched) passive OLE protocol due to Castro et al \cite{Chiraag} that is based on the LWE assumption for generating random OLE. If each server in the main protocol requires $T$ OLE invocations then with a $B$-batched passive OLE protocol we invoke the passive OLE protocol $T/B$ times for each server for a total of $n\cdot T/B$ invocations.
We ran the server protocol between  two Amazon EC2 machines, located in Ohio and N.Virigina. Both machines were  Amazon EC2's ``r4.16xlarge'' instance (2.3 GHz Intel Xeon E5 Broadwell Processors, 488Gb RAM) running Ubuntu 16.08. All our experiments were conducted in two phases. In the first phase, we ran the watchlist setup and generated random OLE  as correlated outputs to the two parties and in the second phase we ran the main protocol to compute the desired functionality, consuming the random OLEs for every OLE invocation in the evaluation.

\begin{figure*}[h]

   	\centering
     \includegraphics[width=8cm]{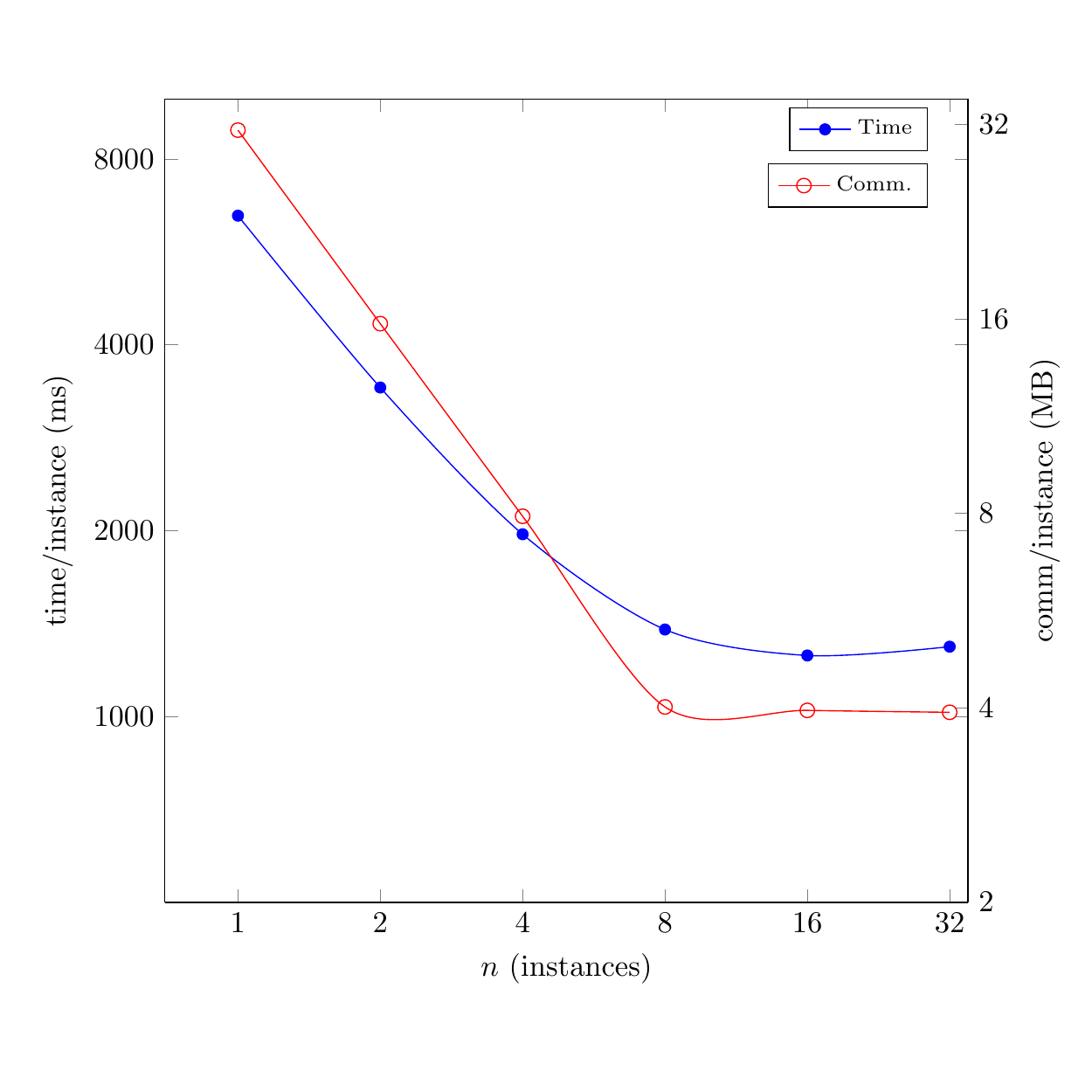}
         \vspace{-4ex}
     \caption{Batched setting in Public NN. }\label{fig:ole}
\end{figure*}

When the NN is assumed to be public, the entire computation (offline and online) took about 3.7s and total communication was 17.89MB. When the NN is private, it took ~259s and the total communication was 3GB. We get added benefits when the computation is highly parallel and therefore we evaluated our system in the batched setting where we carry out the inference on several instances in parallel. The performance is reported in Figure~\ref{fig:ole}. As the instances grows larger, we reach close 1s per instance and 4MB communication per instance when the model is public.  We refer the reader to \cite{MHIV18} for a detailed comparison of the basic secure arithmetic computation protocol with prior works. 


 

%% file: ips.tex
\newcommand{\ol}{\overline{l}}
\newcommand{\lwid}{\ell} 
\newcommand{\lctr}{j} 
\newcommand{\lsctr}{j^*} 
\newcommand{\gctr}{i} 
\newcommand{\gsctr}{i^*} 
\newcommand{\sctr}{c} 
\newcommand{\lval}[2]{L_{#1,#2}} 
\newcommand{\rval}[2]{R_{#1,#2}} 
\newcommand{\outval}[2]{O_{#1,#2}} 
\newcommand{\myrnd}[2]{r_{#1,#2}} 
\newcommand{\myblk}[5]{#4_{#1,#2}^{#5}}
\newcommand{\myblktwo}[4]{#3_{#1,#2}^{#4}}
\newcommand{\lefta}[2]{\myblk{#1}{#2}{v}{a}{L}}
\newcommand{\leftb}[2]{\myblk{#1}{#2}{v}{b}{L}}
\newcommand{\righta}[2]{\myblk{#1}{#2}{v}{a}{R}}
\newcommand{\rightb}[2]{\myblk{#1}{#2}{v}{b}{R}}
\newcommand{\sima}[2]{\myblk{#1}{#2}{v}{a}{O}}
\newcommand{\outa}[2]{\myblk{#1}{#2}{v}{a}{O}}
\newcommand{\outb}[2]{\myblk{#1}{#2}{v}{b}{O}}
\newcommand{\mlt}[2]{\mu_{#1,#2}}
\newcommand{\wmlt}[2]{\widetilde{\mu}_{#1,#2}}

\newcommand{\ua}[1]{u_{#1}}
\newcommand{\ub}[1]{v_{#1}}
\newcommand{\randb}[2]{\myblktwo{#1}{#2}{b}{O}}

\newcommand{\leftsharea}[2]{\myblk{#1}{#2}{s}{\widetilde{a}}{L}}
\newcommand{\leftshareb}[2]{\myblk{#1}{#2}{s}{\widetilde{b}}{L}}
\newcommand{\rightsharea}[2]{\myblk{#1}{#2}{s}{\widetilde{a}}{R}}
\newcommand{\rightshareb}[2]{\myblk{#1}{#2}{s}{\widetilde{b}}{R}}
\newcommand{\outsharea}[2]{\myblk{#1}{#2}{s}{\widetilde{a}}{O}}
\newcommand{\simsharea}[2]{\myblk{#1}{#2}{s}{\widehat{a}}{O}}
\newcommand{\outshareb}[2]{\myblk{#1}{#2}{s}{\widetilde{b}}{O}}
\newcommand{\blindshareb}[2]{\myblk{#1}{#2}{\sigma}{B}{O}}
\newcommand{\uashare}[1]{\widetilde{u}_{#1}}
\newcommand{\ubshare}[1]{\widetilde{v}_{#1}}
\newcommand{\randbshare}[2]{\myblk{#1}{#2}{}{\widetilde{b}}{O}}
\newcommand{\mshare}[2]{m_{#1,#2}}
\newcommand{\outshare}[2]{m_{#1,#2}}

\newcommand{\irs}[1]{\widetilde{\eta}_{#1}}
\newcommand{\perm}[1]{\widetilde{\zeta}_{#1}}
\newcommand{\F}{|\mathcal{F}|}

\section{Secure Arithmetic Computation Protocol of \cite{MHIV18}}\label{app:leviosa}
In this section, we provide the secure arithmetic computation protocol of \cite{MHIV18} (verbatim) for immediate reference. The basic protocol is obtained by modularly combining two ingredients: (1) an outer protocol that is secure in the honest majority setting against active adversaries, and (2) an inner protocol that is secure in the dishonest majority setting against passive adversaries. We describe the outer and inner protocol first. 

\subsection{Our Optimized Outer Protocol}\label{sec:outer}


\begin{figure*}[]
    \caption{\textbf{Optimized Outer Protocol $\Pi$}}
    \label{fig:outer}
    \small
    \begin{framed}

\smallskip\noindent{\bf Inputs.} Client $\alice$'s input is $x=(x_1,\ldots,x_{\alpha_1})$ and client $\bob$'s input is $y=(y_1,\ldots,y_{\alpha_2})$. The parties share a description of an arithmetic circuit $\cir:\FF^{\alpha_1}\times\FF^{\alpha_2}\to \FF$ that implements $\cF$.

\smallskip\noindent{\bf Notations.} $w$ denote the block length, $t$ denote the privacy parameter of the protocol and $e$ is the robustness parameter. We set $k = w + t + e$ is the degree of the polynomials used to shares the blocks and the number servers is $n$.


\smallskip\noindent{\bf The protocol.} The protocol proceeds iteratively starting from the input layer $j=1$ to the last layer $j=d$ 

\smallskip\noindent{\bf Input sharing.} The clients $\alice$ and $\bob$ arrange their inputs values into blocks of length $w$ with a replication pattern that matches the parallel evaluation of the first layer. Namely, for each input block $B_\inp$ of $\alice$ (resp. of $\bob$), it samples a random codeword $u\in L$ such that $B_\inp=\dec_\zeta(u)$, by sampling a random polynomial $p_u(\cdot)$ of degree $<k$ such that $(p_u(\zeta_1),\ldots,p_u(\zeta_w)) = B_\inp$. It then sends $p(\eta_\sctr)$ to server $s_\sctr$. 

\smallskip\noindent{\bf Evaluating the ${\mathbf\lctr^{\mathbf{th}}}$ layer of $\cir$.} Iterating through the layers, the servers do the following for layers $\lctr=1,2,\ldots,d$.
The input wires for the $j^{th}$ layer are organized left and right blocks corresponding to left and right input wires, where the output of each layer can be obtained by performing an arithmetic operation on pairs of input blocks. Furthermore, we assume that each block $B$ corresponds to a vector $u = (u_1,\ldots,u_n)$ such that $u_\sctr$ is held by server $s_\sctr$ and $u$ allegedly encodes the values for the wires within the block. Formally,

\smallskip\noindent {\bf Addition/Subtraction.} Addition or subtraction of the corresponding left and right blocks, respectively denoted by $B_L$ and $B_R$, are handled without any interaction between the servers, by simply having each server locally add or subtract the corresponding shares. Namely, $l_\sctr = u_\sctr + v_\sctr$ where $u$ and $v$ are the vector of shares corresponding to $B_L$ and $B_R$.

\smallskip\noindent {\bf Multiplication.} Multiplication of blocks $B_L$ and $B_R$ is more involved and requires interaction between the clients. Let $u$ and $v$ be the  vector of shares corresponding to these blocks. First, the servers locally multiply their shares $u_\sctr$ and $v_\sctr$ to obtain $l_\sctr$. If $u$ and $v$ belong to $L = \RS_{\FF,n,k,\eta}$, then $l \in L' = \RS_{\FF,n,2\cdot k,\eta}$. The servers then perform a degree reduction to obtain a fresh code word for the same block that lies in $L$.

\smallskip\noindent {\bf Degree reduction.} Since degree reduction can be expressed as a linear transformation on the vector $l$. Namely, there exists a public matrix $A \in \mathbb{F}^{n \times n}$ such that $\ol = A \cdot l$ where if $l \in L'$, then $\ol \in L$ and $\Dec_\zeta(\ol) = \Dec_\zeta(l)$. More precisely, to perform a degree reduction, server $s_\sctr$ computes a 2-out-of-2 additive sharing of $l_\sctr$, namely $l_\sctr^0,l_\sctr^1$ and sends $l_\sctr^0$ to $\alice$ and $l_\sctr^1$ to $\bob$. $\alice$ collects $l^0 = (l_1^0,\ldots,l_n^0)$ and computes ${l'}^0 = A\cdot l^0$. Next, the client samples a random codeword $z$ in $L$ that encodes the all 0s vector and sends $\overline{l}^0_\sctr$ to $\sctr$ where $\overline{l}^0 = (\overline{l}^0_1, \ldots, \overline{l}^0_n) = l'^0+z$. $\bob$ performs an analogous computation and sends $\overline{l}^1$ to the servers. The servers locally add the shares received from the clients.

\smallskip\noindent {\bf Rearranging the blocks for next layer.} Upon completing the computation of some layer, the output blocks are rearranged for the computation in the next layer. This is done the same way as in the above degree reduction, as permuting the elements of a vector that encodes a block can be achieved by a linear transformation. 

\smallskip\noindent{\bf Correctness tests.} See Figure \ref{fig:outer-tests}.

\smallskip\noindent{\bf Output sharing.} Each server sends the clients its shares corresponding to the output blocks. The clients reconstruct the secrets and obtain the final outputs.
\end{framed}
\end{figure*}


\begin{figure*}[]
    \caption{\textbf{Correctness Tests for Protocol $\Pi$}}
    \label{fig:outer-tests}
    \small
    \begin{framed}

After computing all layers, the servers perform the following tests: degree test, permutation test and equality test. Each of these tests are performed $\sigma$ times where $\sigma$ is an input parameter.

\smallskip\noindent{\bf Notations.} A codeword $u\in L$ encodes a block of secrets that sum up to 0 if $\sum_i^w u(\zeta_i)=0$. A codeword $u\in L$ encodes the all 0s block if $u(\zeta_i)=0$ for all $i\in[w]$.

\smallskip\noindent{\bf Degree test.} This test verifies that the vectors corresponding to all input and output blocks of each layer are valid codewords, namely belong to $L$. (We remark that we do not consider the output blocks of the multiplication layers (as they are in $L'$), rather only the output blocks upon performing the degree reduction.) The clients first distribute the vectors $z^0 = (z^0_1,\ldots,z^0_n)$ and $z^1 = (z^1_1,\ldots,z^1_n)$ that allegedly encode random blocks. That is, each server $s_\sctr$ receives the values $z^0_\sctr$ and $z^1_\sctr$ from the respective party $\alice$ and $\bob$. Let $U \in L^m$ denote the matrix that contains the two rows $z^0 = (z^0_1,\ldots,z^0_n)$, $z^1 = (z^1_1,\ldots,z^1_n)$ and the tested blocks $B_1,\ldots,B_{m-2}$. The servers then receive $m$ random field elements $r \in \mathbb{F}^m$ from the coin-tossing oracle $\cF_\coin$ and locally compute $l = r^T U$. That is, each server $s_\sctr$, holding a column $U_\sctr$, locally computes $l_\sctr = r^T U_\sctr$ and broadcasts $l_\sctr$ to all other servers. The servers collect the vector $l = (l_1,\ldots,l_n)$ and abort if $l\not\in L$. Otherwise they proceed to the permutation test.

\smallskip\noindent{\bf Permutation test.} This test verifies that the input blocks of each layer correctly encode the values from the output blocks of the previous layer. Once again, we will consider the matrix $U$ as in the degree test with the exception that the vectors $z^0$ and $z^1$ encode random secrets that sum up to $0$ within each block. Observe that the set of all permutation constraints can be abstracted as a set of public linear constraints $A\in\FF^{mw\times mw}$ and a vector $b$ such that $Ax = b$ where $x$ is the concatenation of the vectors corresponding to the inputs and outputs of each layer throughout the computation that are encoded within $U \in L^m$.

To check this constraint, each server receives a random vector $r\in\mathbb{F}^{mw}$ from the coin-tossing oracle $\cF_\coin$ and computes
$$
r^TA = (r_{11},\ldots,r_{1w},\ldots,r_{m1},\ldots,r_{mw}).
$$
Now, let $r_i(\cdot)$ be the unique polynomial of degree $<w$ such that $r_i(\zeta_\sctr) = r_{i\sctr}$ for every $\sctr \in [w]$ and $i \in [m]$. Then server $s_\sctr$ locally computes $l_\sctr = (r_1(\zeta_\sctr),\ldots, r_m(\zeta_\sctr))^T U_\sctr$ and broadcasts it to the other servers. The servers collect the values and abort if $l = (l_1,\ldots,l_n) \not\in \RS_{\FF,n,k+w,\eta}$ or $x_1+\cdots+x_l \neq 0$ where $x = (x_1,\ldots,x_n) =\dec_\eta(l)$.

\smallskip\noindent{\bf Equality test.} In the equality test, the servers check that the degree reduction was performed correctly. This procedure will be similar to the permutation test but simpler. Namely, each server defines two matrices $U$ and $V$ where $U$ contains the vectors in $(L')^m$ and $V$ contains the vectors after the degree reduction, namely in $L^m$. The servers receive $r \in \FF^{mw}$ from the coin-tossing oracle $\cF_\coin$ and compute the polynomials $r_i(\cdot)$ that encode $(r_{11},\ldots, r_{1w})$ as above. Next, server $s_\sctr$ computes
$$
l_\sctr = (r_1(\zeta_\sctr),\ldots, r_m(\zeta_\sctr))^T U - (r_1(\zeta_\sctr),\ldots, r_m(\zeta_\sctr))^T V
$$
and broadcasts $l_\sctr$. The servers then collect the values and aborts if $l$ does not encode the all 0s block.

\smallskip
\end{framed}
\end{figure*}


In this section we present our optimized outer protocol in the honest majority setting which involves two clients $\alice$ and $\bob$ and $n$ servers. We consider a slight variant of the IPS compiler where we allow the servers in the outer protocol to have access  to a coin-tossing oracle $\cF_\coin$ which upon invocation can broadcast random values to all servers. When compiling this variant, this oracle is implemented via a coin-tossing protocol executed between the clients (cf. Figure \ref{fig:inner}). A crucial ingredient in our construction is the use of  Reed-Solomon codes for computing packed secret shares \cite{FranklinY92}. We start by providing our coding notations and related definitions.

\paragraph{Coding notation.} For a code $C\subseteq \Sigma^n$ and vector $v\in \Sigma^n$, denote by $d(v,C)$ the minimal distance of $v$ from $C$, namely the number of positions in which $v$ differs from the closest codeword in $C$, and by $\Delta(v,C)$ the set of positions in which $v$ differs from such a closest codeword (in case of a tie, take the lexicographically first closest codeword). We further denote by $d(V,C)$ the minimal distance between a vector set $V$ and a code $C$, namely $d(V,C) = \min_{v\in V}\{d(v,C)\}$.

\BD[Reed-Solomon code]
For positive integers $n,k$, finite field $\FF$, and a vector $\eta=(\eta_1,\ldots,\eta_n)\allowbreak\in\FF^n$ of distinct field elements, the code $\RS_{\FF,n,k,\eta}$ is the $[n,k,n-k+1]$ linear code over $\FF$ that consists of all $n$-tuples $(p(\eta_1),\ldots,p(\eta_n))$ where $p$ is a polynomial of degree $<k$ over $\FF$.
\ED

\BD[Encoded message]
Let $L=\RS_{\FF,n,k,\eta}$ be an RS code and $\zeta=(\zeta_1,\ldots,\zeta_w)$ be a sequence of distinct elements of $\FF$ for $w\le k$. For $u\in L$ we define the message $\dec_\zeta(u)$ to be $(p_u(\zeta_1),\ldots,p_u(\zeta_w))$, where $p_u$ is the polynomial (of degree $<k$) corresponding to $u$. For $U\in L^m$ with rows $u^1,\ldots,u^m\in L$, we let $\dec_\zeta(U)$ be the length $mw$ vector $x = (x_{11},\ldots,x_{1w},\ldots,x_{m1},\allowbreak\ldots,x_{mw})$ such that $(x_{i1},\ldots,x_{iw})=\dec_\zeta(u^i)$ for $i\in [m]$.
We say that {\em $u$ encodes $x$} if $x=\dec_\zeta(u)$.
\ED

We further denote by replication pattern the replication induced by structure of circuit $\cir$ between every pair of layers. A formal description of our protocol is given in Figures \ref{fig:outer}, \ref{fig:outer-tests}.

\medskip\noindent
We now have the following theorem for our protocol.

\medskip
\BT\label{thm:outer}
Let $k > t+e+w$ and let $\cF:\FF^{\alpha_1}\times\FF^{\alpha_2}\to \FF$ be a two-party functionality, then protocol $\Pi$ from Figure \ref{fig:outer} securely computes $\cF$ tolerating static active corruption of at most one client and adaptive active corruptions of at most $e$ servers and passive corruptions of at most $t$ servers with statistical security of $(e+2)/|\FF|^\sigma$ (where $\sigma$ is a soundness amplification parameter).
\ET

\subsection{The Inner Protocol}

Recall that in the  IPS compiler, the inner protocol is a two-party protocol executed between the two parties $\Alice$ and $\Bob$ and security is required to hold only against corruption by a passive adversary. Furthermore, the functionalities considered are precisely the next message functions executed by the servers in the outer protocol. On a high-level, the state of each of server is maintained jointly by the parties where each holds a share. Emulating the internal computation of each server for our outer protocol boils down to securely updating the states of the servers based on the computation specified in Figures~\ref{fig:outer} and \ref{fig:outer-tests}. We remark that all computations performed by the servers are arithmetic computations over the same field. For the inner protocol and we will rely on the GMW protocol \cite{GoldreichMW87} described in the (passive) OLE-hybrid, where the OLE functionality can be instantiated with \emph{any} passively secure protocol \cite{NaorP99,IshaiPS09}.

%% file: inner.tex

\subsection{The Combined Protocol}\label{sec:inner}

In this section we provide our complete two party protocol for realizing arithmetic functions over any field $\FF$ that achieve security against active corruptions. This is obtained by compiling our outer protocol described in \ref{sec:outer} and the inner protocol instantiated using the GMW protocol \cite{GoldreichMW87} with a variant of the IPS compiler. 
We recall that, in IPS compiler, the parties executing the protocol will play the respecitve roles in the inner protocol, and play the roles of the clients in the outer protocol. The actual server actions are never executed by a single party. Instead, the state of the servers are shared using an additive secret sharing scheme between the parties and each server computation is emulated jointly by the two parties via the inner protocol to update the state. Finally, an important part of the compiler is the watchist mechanism where each party monitors the actions of the other party when emulating the servers. Roughly speaking, each party knows the inputs and randomness used by the other party to emulate the server for some hidden subset of the servers refered to as as the watchlist. A party is caught if it deviates in executing the inner protocol corresponding to a server that is on the other party's watchlist. 

A formal description of the protocol is given in Figure \ref{fig:inner}.


\protocoll{The Combined Protocol $\Phi$}{Arithmetic 2PC with active security}{fig:inner}{

\noindent{\bf Inputs.} $\Alice$'s input is $x=(x_1,\ldots,x_{\alpha_1})$ and $\Bob$'s input is $y=(y_1,\ldots,y_{\alpha_2})$. The parties share a description of an arithmetic circuit $\cir:\FF^{\alpha_1}\times\FF^{\alpha_2}\to\FF$ that implements $\cF$.

\smallskip\noindent{\bf Watchlists setup.} To establish the watchlist, $\Alice$ and $\Bob$ run two instances of an actively secure $t$-out-of-$n$ oblivious-transfer (OT) protocol where $t$ is the privacy parameter of the outer protocol. In one instance $\Alice$ plays the role of the sender with $n$ symmetric keys $(k_1,\ldots,k_n)$ as input and $\Bob$ plays the receiver with an arbitrary $t$-subset of $[n]$ as its watchlist. In the second instance, the parties execute the same protocol with the roles reversed. The symmetric keys will be used to encrypt the input and randomness used by the party for executing the inner protocol corresponding to the servers where the information for the $i^{th}$ server is encrypted with the $i^{th}$ key. This way each party learns the information corresponding to the servers  on their watchlists and monitors the behaviour of the other party. 


\smallskip\noindent{\bf Input sharing.} For each $i\in[m]$, $\Alice$ secret shares its input using an additively secret sharing scheme. Namely, $\Alice$ picks $n$ pairs of random field elements $((x_1^1,x_1^2),\ldots,(x_m^1,\allowbreak x_m^2))$ subject to $x_i = x_i^1+x_i^2$ for all $i\in[m]$, and sends $(x_1^2,\ldots x_m^2)$ to $\Bob$. Similarly, $\Bob$ secret shares its input and sends the shares $(y_1^1,\ldots y_m^1)$ to $\Alice$.

\smallskip\noindent{\bf Protocol emulation.} The parties proceed iteratively starting from the input layer $j=1$ to the last layer $j=d$, emulating the actions of the clients and servers in Protocol $\Pi$. In more details, the parties play the roles of the clients exactly as carried out in $\Pi$. To emulate the computations of the servers, the parties maintain throughput the emulation of $\Pi$ additive shares of the local state of each server, that contains the incoming and outgoing messages until that point. Then, additions are emulated locally by adding the corresponding shares. Next, to emulate the multiplications carried out by each server, the parties employ the inner protocol to multiply from input shares of the elements to receive shares of the product. This is the GMW protocol that makes use of two OLE calls. Finally, to emulate the communication from $\client$ to $s_j$, $\party$ encrypts the message to be sent from $\client$ to $s_j$ using secret key $k_j$ and sends the ciphertext to the other party. To simulate a server communicates a message to a client, the other client simply reveals it share of the message being communicated.

\smallskip\noindent{\bf Coin tossing.} Whenever the outer protocol invokes the coin-tossing oracle $\cF_\coin$, the parties run a coin tossing protocol using the commitment functionality $\cF_\comm$, for producing the randomness that is used as the randomness obtained from the coin tossing oracle for the outer protocol. Namely, each party commits to each share, upon both committing the parties open their commitments and XOR the shares.

\smallskip\noindent{\bf Checking consistency.} Each party $\party$ who has server $s_j$ in its watchlist $\watch_j$ carries out a consistency check between the messages reported over the watchlist channel and the messages it receives from $s_j$. Specifically, party $\party$ recomputes the messages sent by $s_j$ using the messages reported on the watchlist channel $\watch_j$, and aborts in case of detecting inconsistency.

}

\noindent We conclude with the following theorem.

\smallskip
\BT\label{thm:main}
Let $\cF:\FF^{\alpha_1}\times\FF^{\alpha_2}\to \FF$ be a two-party functionality with depth $d$. Let $\Pi$ be the outer MPC protocol described in Section~\ref{sec:outer} for realizing $\cF$ with $n$ servers, $e$-robustness, $(t+e)$-privacy, $\sigma$ repetitions, $w$ block length and $L=RS_{\FF,n,k,\eta}$. Let $g$ be the multiplication functionality that takes additive shares as inputs from the parties and outputs additive shares of the product and $\rho^\ole$ a two-party protocol that realizes $g$ in the passive OLE-hybrid setting with perfect security against a passive corruption of either party. Then, if $k > t+e+w$ and $e < (n-k)/3$, the compiled protocol from Figure \ref{fig:inner}, denoted by $\Phi_{\Pi,\rho}^{\ole,\ot_{t,n},\comm}$, securely realizes $\cF$ in the passive OLE-hybrid model against active (static) corruption of either party with statistical security
$$
(e+2)/|\FF|^\sigma+ (1-e/n)^t + ((3\cdot e+2\cdot w+2\cdot t)/n)^t.
$$
The communication complexity of $\Phi$ is
\begin{align*}
&\underbrace{2\cdot \ccc_{\mbox{\rm t-out-of-n OT}}}_{\mbox{\rm{watchlist setup}}}~
+ \underbrace{n\cdot d\cdot \ccc_\rho}_{\mbox{\rm{passive invoc.}}}
+ \underbrace{d\cdot n \cdot \log_2|\FF|}_{\mbox{\rm{output in each layer}}} \\
& + \underbrace{3\cdot\kappa}_{\mbox{\rm{coin-toss}}}
+ \underbrace{2\cdot \sigma\cdot (t+e+w)\cdot \log_2|\FF|}_{\mbox{\rm{degree test}}} \\
& + \underbrace{4\cdot \sigma\cdot (t+e+w)\cdot \log_2|\FF|}_{\mbox{\rm{perm./equality test}}}
\end{align*}
where $\ccc_\rho$ is the communication complexity of $\rho^{\ole}$ and $\sigma$ is a soundness amplification parameter. In addition, the number of OLE invocations of $\Phi$ is $n\cdot d \cdot r^\rho_{\ole}$ where $r^\rho_{\ole}$ is the number of OLE invocations of $\rho^{\ole}$.
\ET